\documentclass[12pt]{article}
\usepackage{graphicx}
\catcode`\@=11

\global\arraycolsep=2pt
\usepackage{amsbsy,amssymb,latexsym,amsfonts,amsmath}
\usepackage{graphicx,color}

\begin{document}
\begin{flushright}
{RUP-17-8}
\end{flushright}

\vspace*{0.7cm}

\begin{center}
{ \Large Bootstrap experiments on higher dimensional CFTs \\}
\end{center}
\vspace*{1.0cm}
\begin{center}
{Yu Nakayama}
\vspace*{1.0cm}

Department of Physics, Rikkyo University, Toshima, Tokyo 171-8501, Japan

and 

Kavli Institute for the Physics and Mathematics of the Universe (WPI),  
\\ University of Tokyo, 5-1-5 Kashiwanoha, Kashiwa, Chiba 277-8583, Japan

\vspace{3.8cm}
\end{center}

\begin{abstract}
Recent programs on conformal bootstrap suggest an empirical relationship between the existence of non-trivial conformal field theories and non-trivial features such as a kink in the unitarity bound of conformal dimensions in the conformal bootstrap equations. We report the existene of non-trivial kinks in the unitarity bound of scalar operators in the adjoint representation of the $SU(N)$ symmetric conformal field theories. They have interesting properties (1) the kinks exist in $d<6$ dimensions (2) the location of kinks are when the unitarity bound hits the space-time dimension $d$ (3) there exists a ``conformal window" of $N<N_*$, where $N_* \sim 15 $ in $d=4$ and $N_* \sim 20$ in $d=5$. 


\end{abstract}

\thispagestyle{empty} 

\setcounter{page}{0}

\newpage
\section{Introduction}
Finding evidence for the existence of non-trivial conformal field theories beyond perturbation theory is of importance from various viewpoints. Such theories in $d=3$ or $d=4$ space-time dimensions may have potential applications to condensed matter physics and high energy particle physics. Theoretically it is even more challenging to find such in $d>4$ dimensions because it is difficult to find any reasonable Lagrangian descriptions with perturbative renormalizability, and it is believed that we need non-perturbative tools to address them.

In recent years, we have developed a completely non-perturbative approach to the very  (non-)existence of conformal field theories based on the numerical conformal bootstrap \cite{Rattazzi:2008pe}-\cite{Simmons-Duffin:2015qma}. Translating the conformal bootstrap equations with a given global symmetry to a certain optimization program (e.g. semi-definite program), one may give a rigorous bound on the conformal data, i.e. conformal dimensions and operator product expansion (OPE) coefficients, of unitary conformal field theories. 

A priori, the constraint on the conformal data from the conformal bootstrap equations has little to say about  the existence (rather than the non-existence) of conformal field theories under investigation. However, there is surprising empirical evidence that a non-trivial feature in the unitarity bound of the conformal data may indicate the existence of actual conformal field theories. There is no theoretical proof of such a statement, but we have seen many examples such as Ising model, $O(N)$ vector models  and supersymmetric theories. Applying this idea gives strong motivations to  identify such theories and establish the``conformal window" associated with the feature.  

In this paper, we further pursue this experimental search for non-trivial features in unitarity bound of conformal data from the conformal bootstrap equations.  We report the existence of non-trivial kinks in the unitarity bound of the scalar operators in the adjoint representation of $SU(N)$ symmetric conformal field theories. They have interesting properties (1) they exist in $d<6$ dimensions (2) the location of kinks are when the unitarity bound hits the space-time dimension $d$ (3) there exists a ``conformal window" of $N<N_*$, where $N_* \sim 15 $ in $d=4$ and $N_* \sim 20$ in $d=5$. 

In the simplest implementation of the conformal bootstrap equation with the only $\mathbb{Z}_2$ symmetry (aiming at the Wilson-Fisher fixed point), the upper critical dimension was $d=4$ from the existence of features in the unitariy bound of the conformal bootstrap equation \cite{El-Showk:2013nia}, which coincides with the upper critical dimension of the Wilson-Fisher fixed point of the Landau-Ginzburg model. Our finding of upper critical dimension of $d=6$ is much more mysterious and suggestive. It is widely believed (with no rigorous justification) that there does not exist any non-trivial conformal field theory in $d>6$, and our finding is consistent with such a claim.

The organization of this paper is as follows. In section 2, we present the method of the numerical conformal bootstrap program. In section 3, we present our results with many figures. In section 4, we present the discussions. In appendix, we show the results of the numerical conformal bootstrap program in the other sectors.

\section{Method}
In this paper, we work on the $d$-dimensional conformal field theories with $SU(N)$ global symmetry, and study the consistency of crossing equations for the four-point functions among scalar primary operators $\Phi$ in the adjoint representation. Decomposing the OPE of $\Phi\times \Phi$ into the irreducible representations,\footnote{In terms of the Young Tableau, we have $\mathrm{Adj} = [N-1,1]$, $(\mathrm{A\bar{S}}) = [N-1,N-1,2]$, $(\mathrm{A\bar{A}})= [N-2,2]$, and $(\mathrm{S\bar{S}}) = [N-1,N-1,1,1]$, where the number in the bracket denotes the number of column boxes.} we obtain the OPE sum rule for the four-point functions or the conformal bootstrap equations:
\begin{align}
0 =& \sum_{O \in \Phi \times \Phi} \lambda_O^2 V^{(+)}_{1}+ \sum_{O \in \Phi \times \Phi} \lambda_O^2 V^{(-)}_{\mathrm{Adj}} +\sum_{O \in \Phi \times \Phi}  \lambda_O^2 V^{(+)}_{\mathrm{Adj}} \cr
 &+  \sum_{O \in \Phi \times \Phi} \lambda_O^2 V^{(-)}_{(\mathrm{S\bar{A}}) +cc} +  \sum_{O \in \Phi \times \Phi} \lambda_O^2 V^{(+)}_{(\mathrm{A\bar{A}})} +  \sum_{O \in \Phi \times \Phi} \lambda_O^2 V^{(+)}_{(\mathrm{S\bar{S}})}
\end{align}
where $(\pm)$ denotes the even $(+)$ or odd $(-)$ spin contributions. By using the convention
\begin{align}
F & = v^{\Delta_{\Phi}} g_{\Delta_O,l}(u,v) - u^{\Delta_{\Phi}} g_{\Delta_O,l}(v,u) \cr
H & = v^{\Delta_{\Phi}} g_{\Delta_O,l}(u,v) + u^{\Delta_{\Phi}} g_{\Delta_O,l}(v,u)
\end{align}
with the conformal block $g_{\Delta_O,l}$ being normalized as in \cite{Hogervorst:2013kva}, whose explicit expression can be found in \cite{Dolan:2003hv},
the each representation contributes to the sum rule as
\begin{align}
V_{1}^{(+)} &= \left( \begin{array}{cc}  0 \\ 0 \\ 0 \\ F \\ H \\  0 \\
\end{array} \right) \ , \ \
V_{\mathrm{Adj}}^{(-)} = \left( \begin{array}{cc}  0 \\ 0 \\ -F \\ 0 \\ 0\\  -H \\
\end{array} \right) \ , \ \
V_{\mathrm{Adj}}^{(+)} = \left( \begin{array}{cc}  0 \\ 2N^{-1}F \\ -F \\ -16N^{-1}F \\ -4N^{-1} H \\  H \\
\end{array} \right) \ , \cr
V^{(-)}_{(\mathrm{S\bar{A}})+cc} &=  \left( \begin{array}{cc}  -F \\ 0 \\ N^{-1}F \\ 0 \\ - H \\  N^{-1}H \\
\end{array} \right) \ , \ \
V_{\mathrm{A\bar{A}}}^{(+)} = \left( \begin{array}{cc} F \\ -(N-2)^{-1} F \\  (N-2)^{-1} F   \\ \frac{2N^2}{N^2-3N+2} F \\ -\frac{N(N-3)}{N^2-3N+2} H \\ \frac{N-3}{N-2}H \\ 
\end{array} \right) \ , \ \
V^{(+)}_{(\mathrm{S\bar{S}})}  \left( \begin{array}{cc} F \\ (N+2)^{-1} F \\  (N+2)^{-1} F \\ \frac{2N^2}{N^2+3N+2} F   \\ -\frac{N(N+3)}{N^2+3N+2} H \\ -\frac{N+3}{N+2}H \\
\end{array} \right) \ . \ \
\end{align}
 We note that for this four-point function, the sum rule is the same with or without extra $U(1)$. In other words, our bounds apply not only to the $SU(N)$ symmetric theories but also $U(N)$ symmetric theories.

This equation has been derived in \cite{Berkooz:2014yda}\cite{Iha:2016ppj} with  a  different focus in mind. The former is interested in the supersymmetric field theories and the latter is interested in its application to many flavor conformal QCD in $d=4$.
In this paper, we focus on the unitarity bound of the conformal dimensions of scalar operators in the adjoint representation that appear in the $\Phi \times \Phi$ OPE. Note that from the group theory perspective, it is always possible that $\Phi$ itself appears in the OPE of $\Phi \times \Phi$, but our numerical analysis does not assume it, and our bound allows that the lowest dimensional operator in the adjoint representation that appear in the $\Phi \times \Phi$ OPE may have larger conformal dimension than that of $\Phi$. In such cases, we effectively impose the extra $\mathbb{Z}_2$ symmetry acting on $\Phi$ (that maps $\Phi$ to $-\Phi$) so that $\Phi$ itself does not appear in the OPE. 

Our implementation of the numerical conformal bootstrap is based on cboot \cite{cboot}, which uses the SDPB \cite{Simmons-Duffin:2015qma} as a core of the semi-definite programming. For the truncation of the search space by number of derivatives, we use $\Lambda (= N_{\mathrm{max}}) = 23$. The other parameters such as the number of included spin are chosen appropriately so that the numerical  optimization is stable.\footnote{All the numerical computations in this paper are done on a single 8-core desktop computer.}
The numerical bootstrap program is mostly analytic in space-time dimensions $d$, so one may change the space-time dimensions freely.\footnote{The only practical exception is even integer space-time dimensions, where the rational approximation of the conformal blocks become complicated due to the double poles. We give a special treatment in $d=4$ by using the explicit form of given by hypergeometric functions. When we say $d=6$ in this paper, we actually work in  $d=5.999$ dimensions.}

We will see non-trivial features in the unitarity bound of the conformal dimensions of scalar operators in the adjoint representation. With the empirical working hypothesis of ``kinks $=$ conformal field theories", we try to identify them as non-trivial conformal field theories with the $SU(N)$ symmetry. We, in particular, pay attention when the kinks disappear by changing $d$ and $N$, which gives a ``conformal window" of the unitary conformal field theories under investigation.

In addition to the unitarity bound on the conformal data, the numerical conformal bootstrap program is able to tell us much more information of the putative CFTs realized at the boundary of the unitarity bound. 
From the consideration of how the optimization problem works in the numerical bootstrap program, we expect that the conformal data that saturates the unitarity bound obtained as an output of the semi-definite program is unique. If this is the case, we may determine the conformal data that saturates the unitarity bound from the optimization \cite{ElShowk:2012hu}.
This uniqueness indeed holds a posteori in our numerical conformal bootstrap program, so we are able to read the entire spectrum of the putative conformal field theories realized at the boundaries of the unitarity bound, especially at the kinks. We will report these data in the next section.

\section{Results}
\begin{figure}[htbp]
	\begin{center}
		\includegraphics[width=12.0cm,clip]{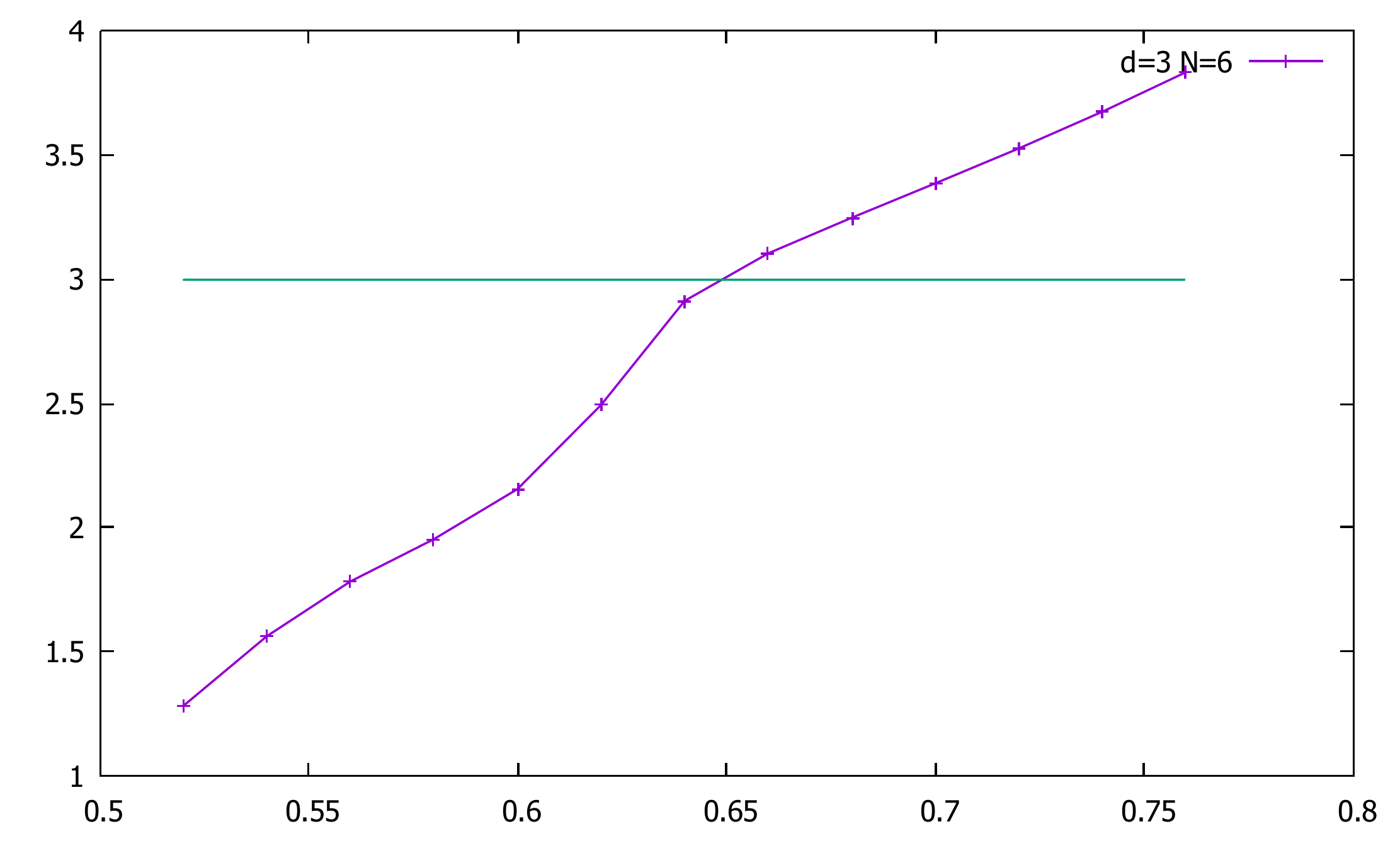}
	\end{center}
	\caption{Unitarity bound in $d=3$ with $N=6$ in the adjoint sector.}
	\label{fig:d3n6}
\end{figure}

\begin{figure}[htbp]
	\begin{center}
		\includegraphics[width=12.0cm,clip]{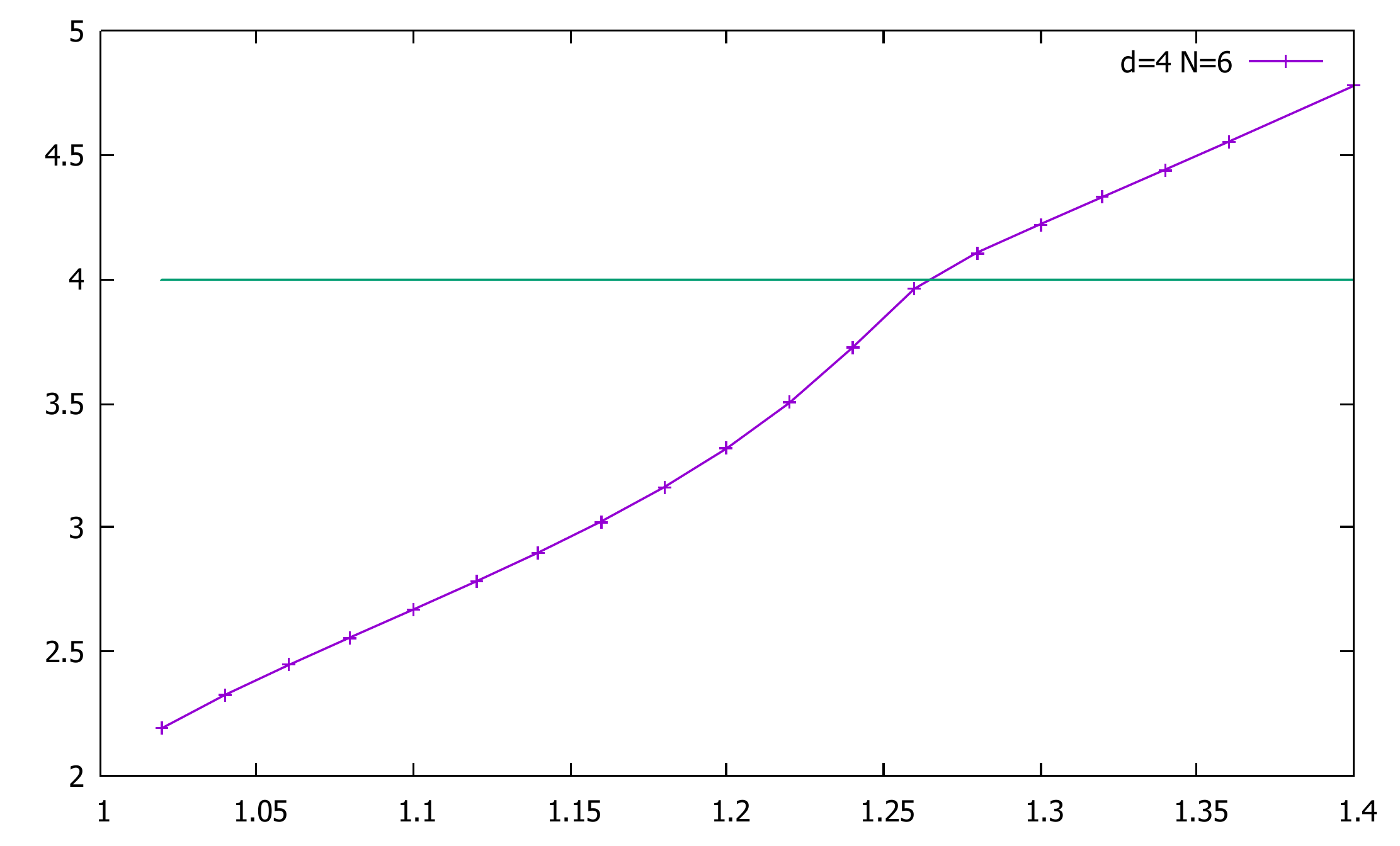}
	\end{center}
	\caption{Unitarity bound in $d=4$ with $N=6$ in the adjoint sector.}
	\label{fig:d4n6}
\end{figure}

\begin{figure}[htbp]
	\begin{center}
		\includegraphics[width=12.0cm,clip]{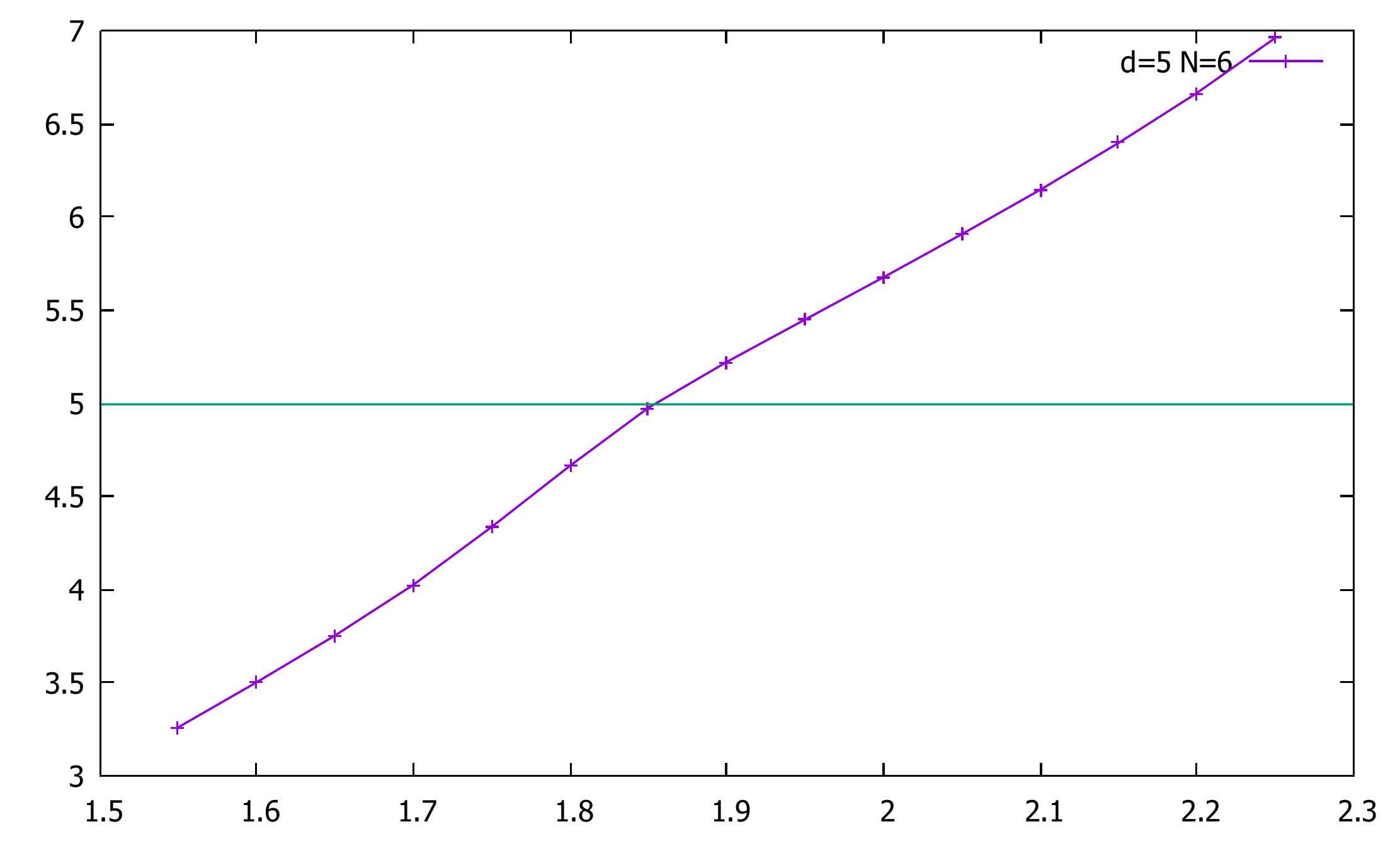}
	\end{center}
	\caption{Unitarity bound in $d=5$ with $N=6$ in the adjoint sector.}
	\label{fig:d5n6}
\end{figure}

\begin{figure}[htbp]
	\begin{center}
		\includegraphics[width=12.0cm,clip]{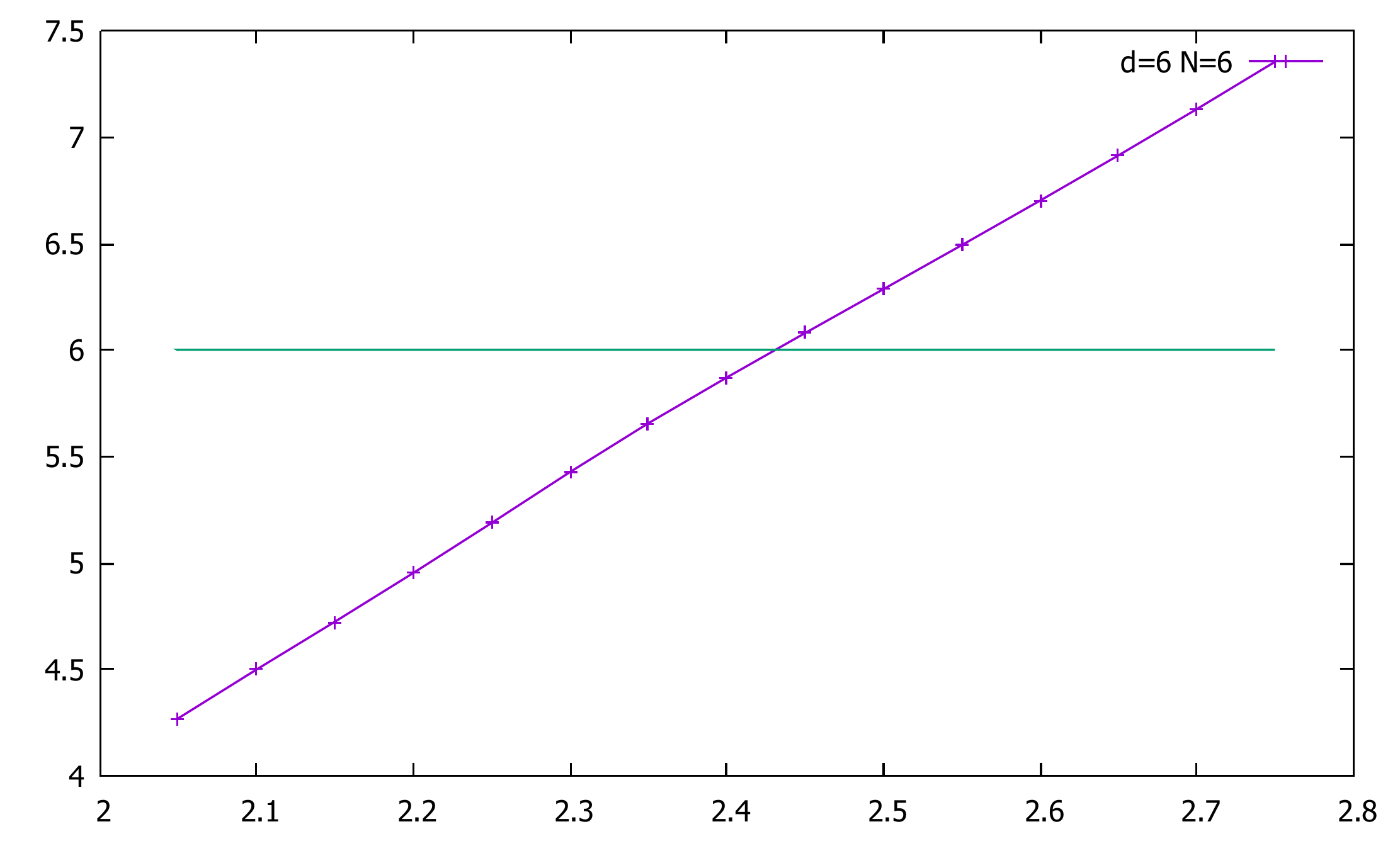}
	\end{center}
	\caption{Unitarity bound in $d=6$ with $N=6$ in the adjoint sector.}
	\label{fig:d6n6}
\end{figure}

\begin{figure}[htbp]
	\begin{center}
		\includegraphics[width=12.0cm,clip]{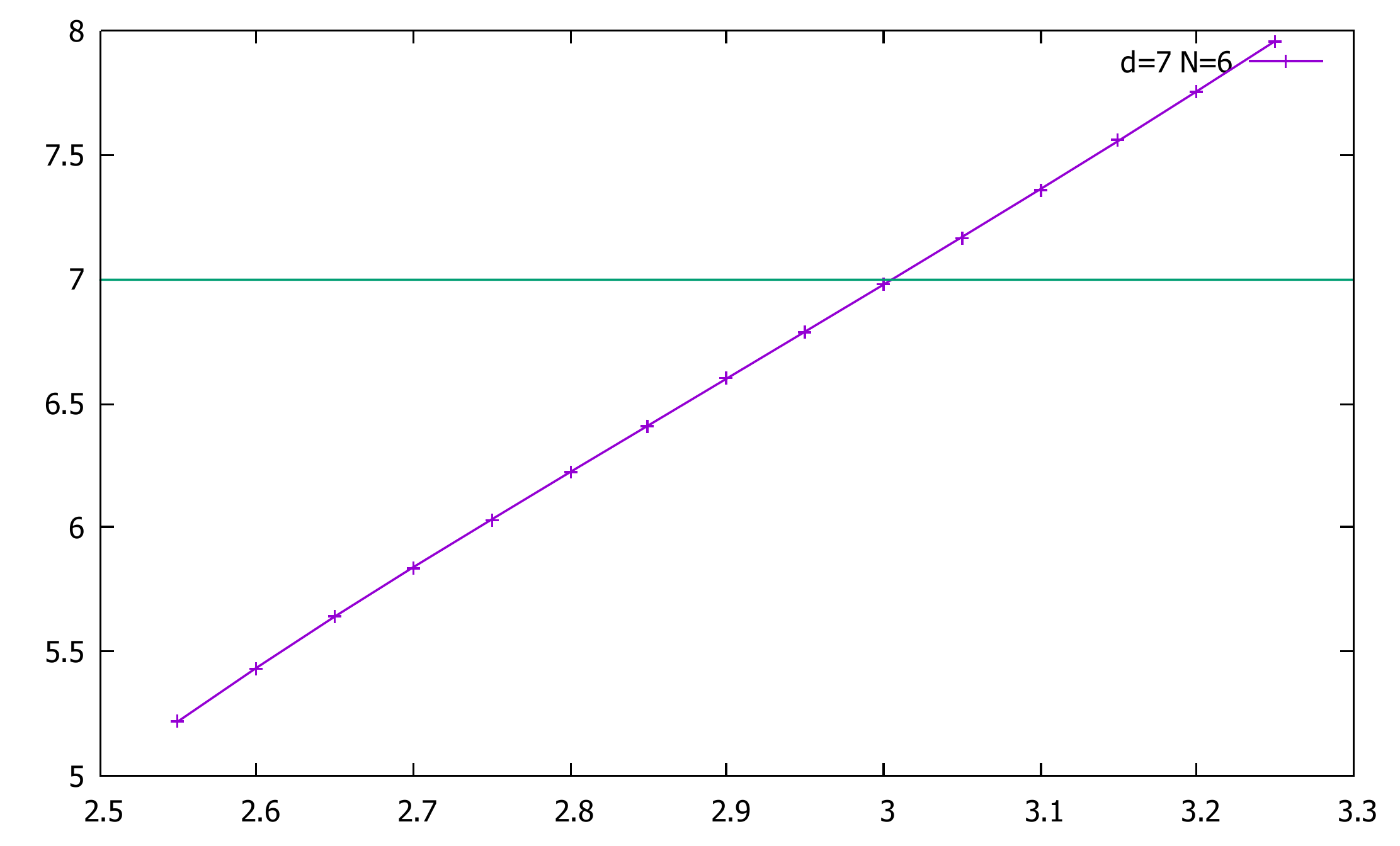}
	\end{center}
	\caption{Unitarity bound in $d=7$ with $N=6$ in the adjoint sector.}
	\label{fig:d7n6}
\end{figure}

In Fig \ref{fig:d3n6}-\ref{fig:d7n6} we first show the unitarity bound of the conformal dimensions of the scalar operators in the adjoint representation of $SU(N)$ symmetric conformal field theories with $N=6$ in various dimensions $d=3,4,5,6,7$. The horizontal axis is the conformal dimension of a primary scalar operator $\Phi$ in the adjoint representation, and the vertical axis is the bound  of  the conformal dimension of the primary scalar operator that is in the adjoint representation and that appears in the $\Phi \times\Phi$ OPE.
The bound means that the conformal data is consistent below the curve presented in these figures. 
We have chosen $N=6$ because the feature looks the most eminent than the other $N$. We discuss this point below.

We see that in $d=3,4,5$ we have non-trivial (kink-like) features when the unitarity bound of the scalar operator in the adjoint representation (approximately) hits the space-time dimension $d$. The feature looks weaker in $d=5$, but it is still visible. In $d=6$ and $d=7$, the feature seems to disappear completely. If we continue the space-time dimensions to a real number, which is possible in numerical conformal bootstrap program, then we may see that the feature continuously becomes weaker toward $d=6$ dimensions.

It is not obivous to us what is the underlying reason to have kinks when the unitarity bound hits the space-time dimension $d$. Certainly when the conformal dimension is equal to the space-time dimension, we have an extra marginal deformation, and such CFTs, if any, are special, but we do not know the precise reason why such a physical condition plays a role in the conformal bootstrap equations. At the same time, we note that we have also observed non-trivial features (i.e. spikes there) that appear when the bound hits the space-time dimension $d$ in the mixed correlator conformal bootstrap with $O(2)$ symmetry in \cite{Nakayama:2016jhq}.

\begin{figure}[htbp]
	\begin{center}
		\includegraphics[width=12.0cm,clip]{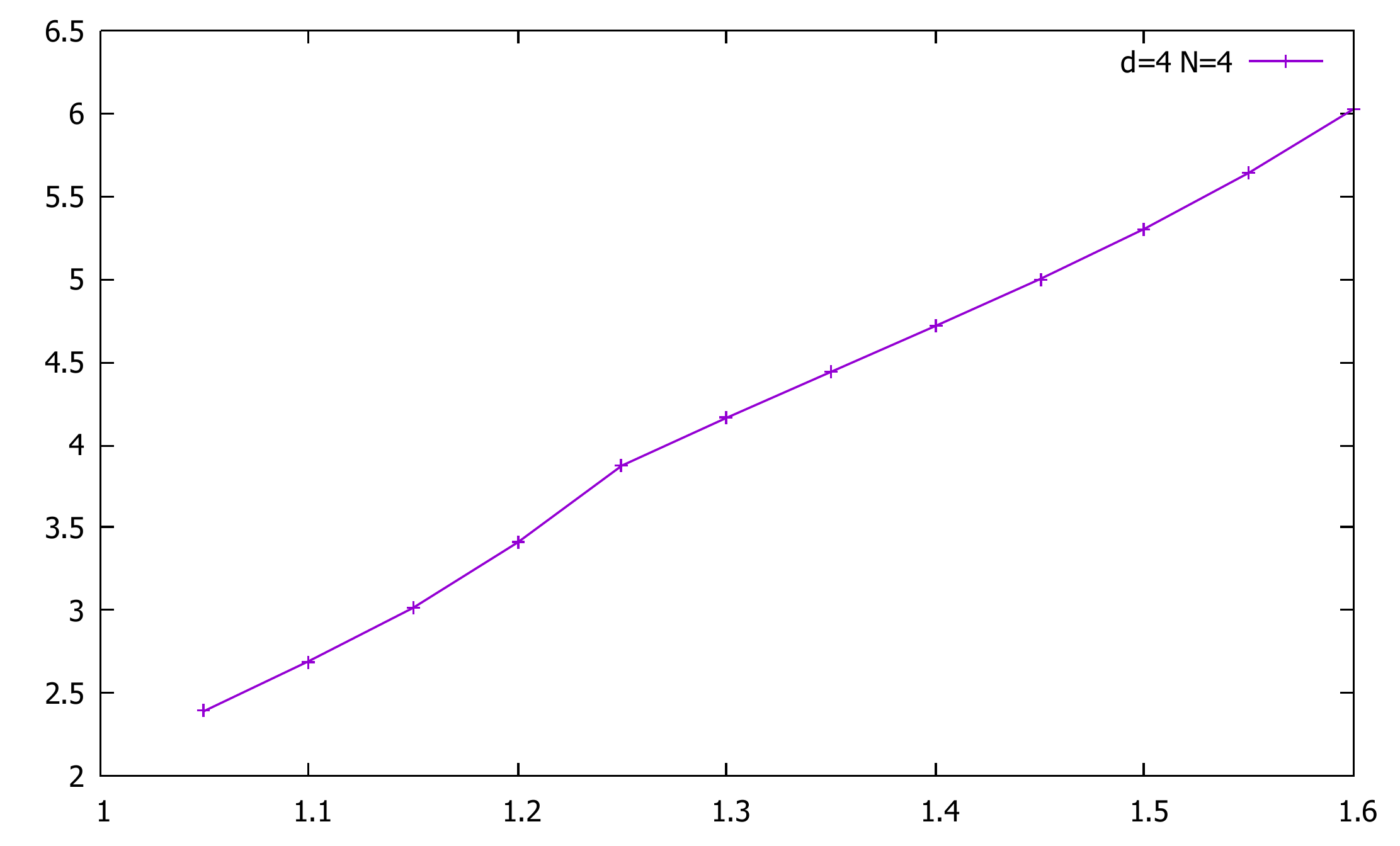}
	\end{center}
	\caption{Unitarity bound in $d=4$ with $N=4$ in the adjoint sector.}
	\label{fig:d4n4}
\end{figure}

\begin{figure}[htbp]
	\begin{center}
		\includegraphics[width=12.0cm,clip]{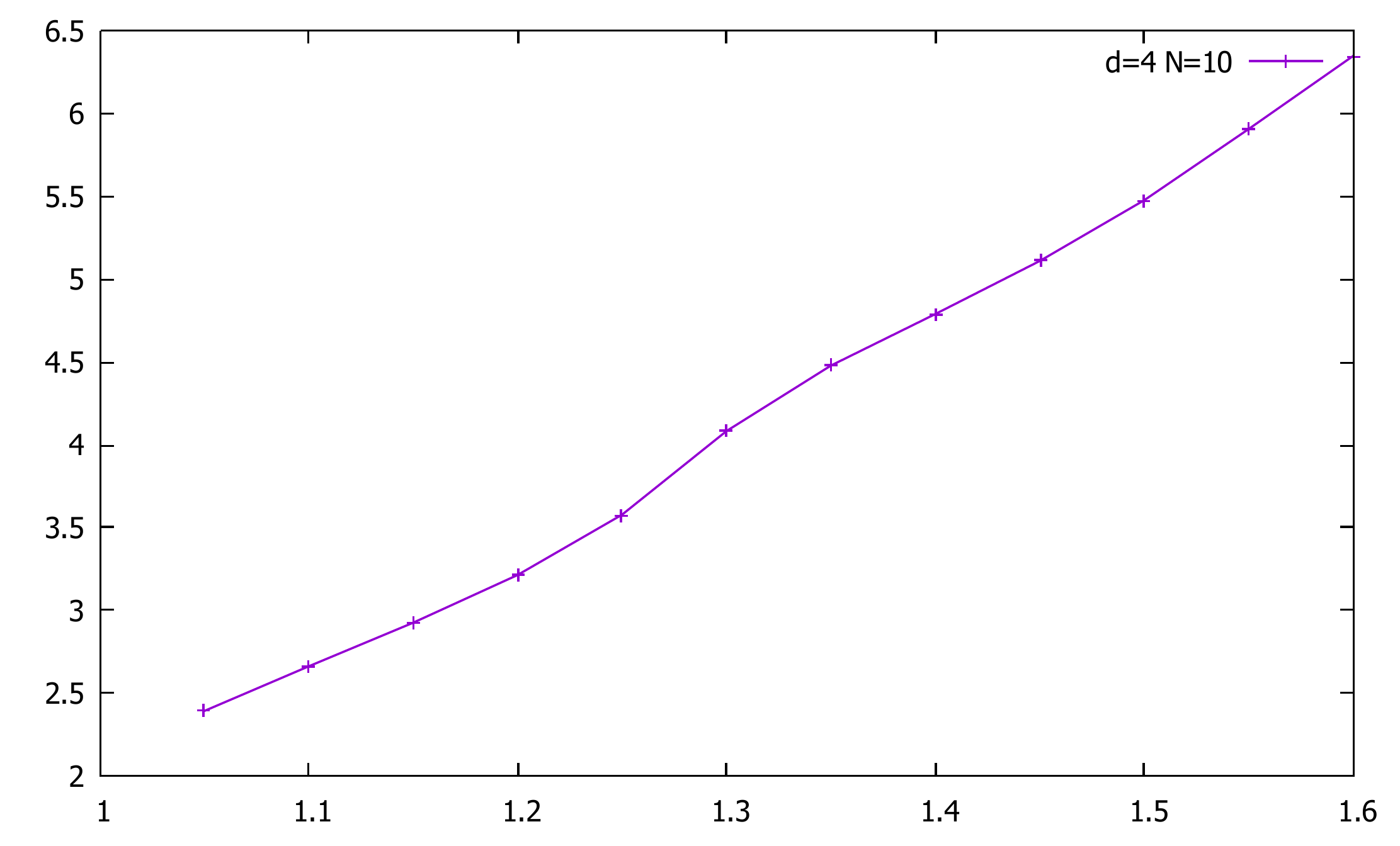}
	\end{center}
	\caption{Unitarity bound in $d=4$ with $N=10$ in the adjoint sector.}
	\label{fig:d4n10}
\end{figure}

\begin{figure}[htbp]
	\begin{center}
		\includegraphics[width=12.0cm,clip]{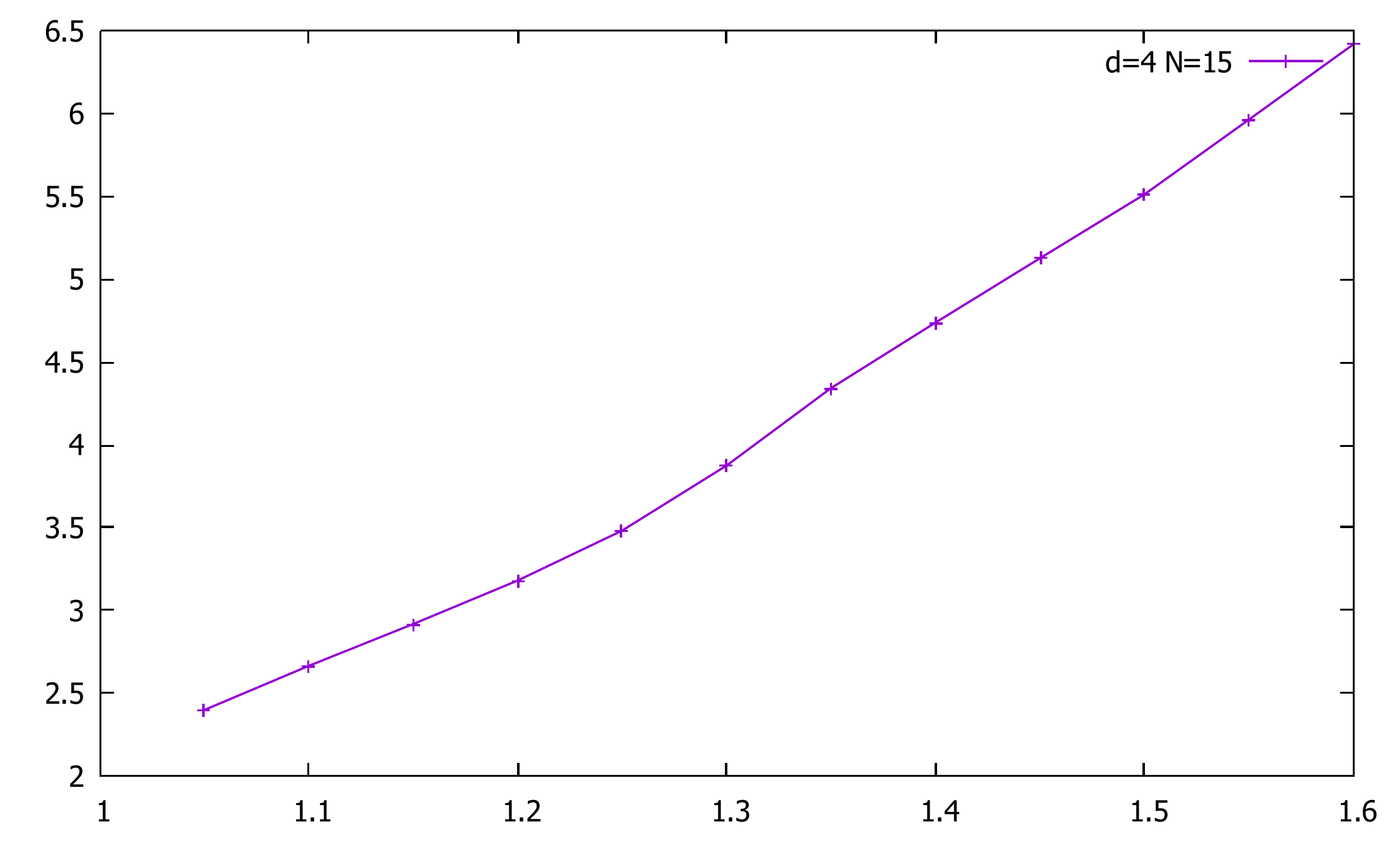}
	\end{center}
	\caption{Unitarity bound in $d=4$ with $N=15$ in the adjoint sector.}
	\label{fig:d4n15}
\end{figure}

\begin{figure}[htbp]
	\begin{center}
		\includegraphics[width=12.0cm,clip]{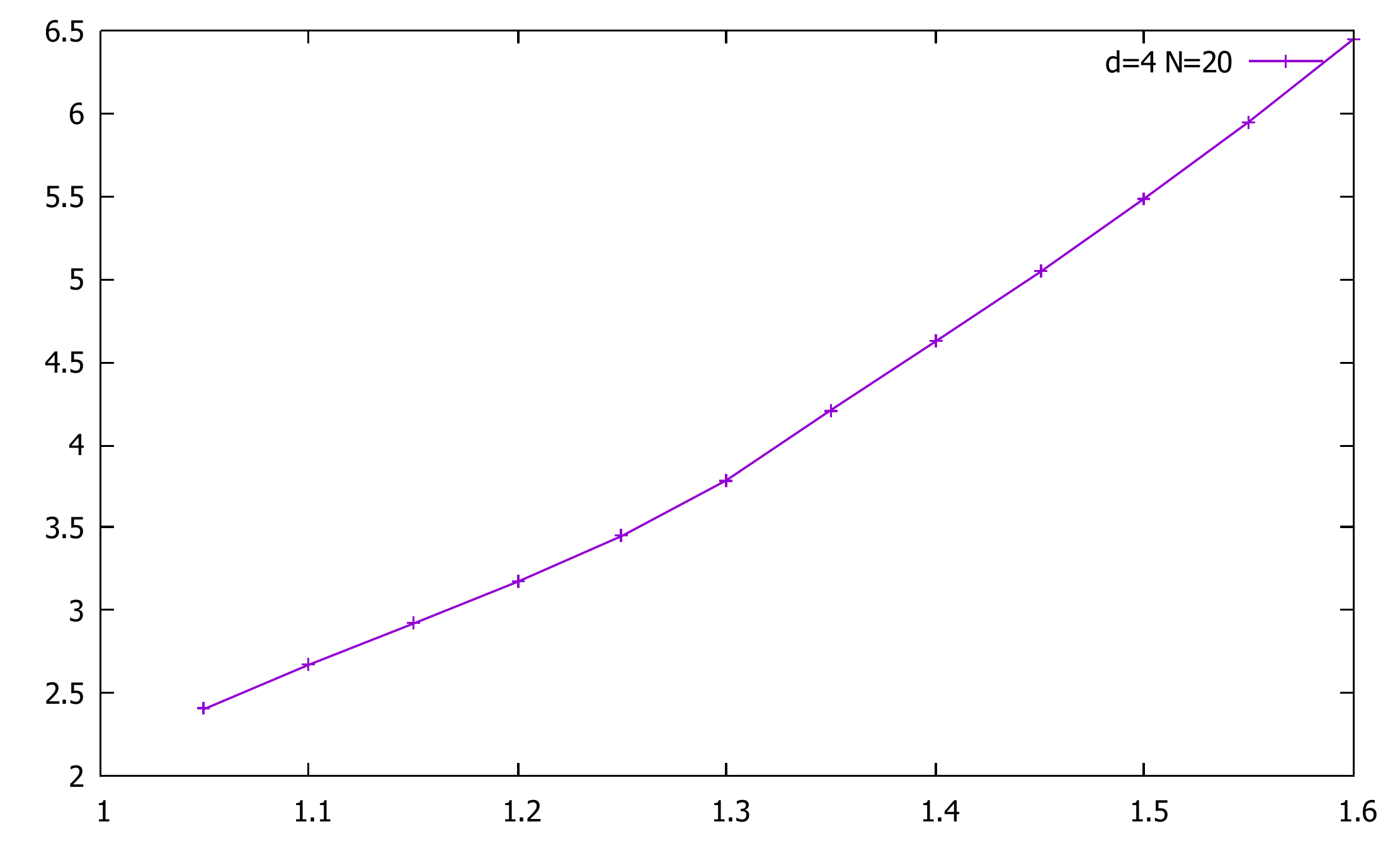}
	\end{center}
	\caption{Unitarity bound in $d=4$ with $N=20$ in the adjoint sector.}
	\label{fig:d4n20}
\end{figure}

\begin{figure}[htbp]
	\begin{center}
		\includegraphics[width=12.0cm,clip]{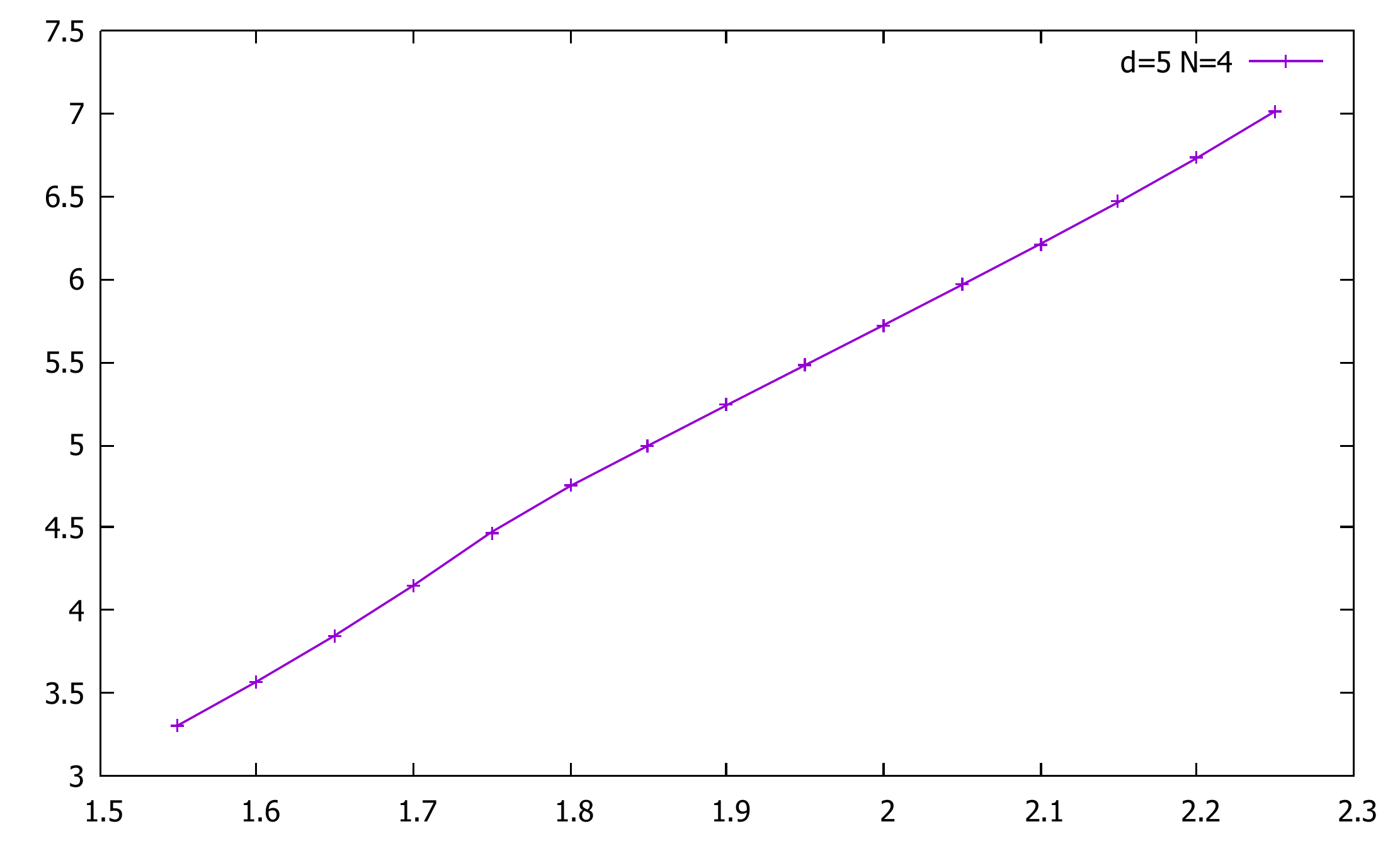}
	\end{center}
	\caption{Unitarity bound in $d=5$ with $N=4$ in the adjoint sector.}
	\label{fig:d5n4}
\end{figure}

\begin{figure}[htbp]
	\begin{center}
		\includegraphics[width=12.0cm,clip]{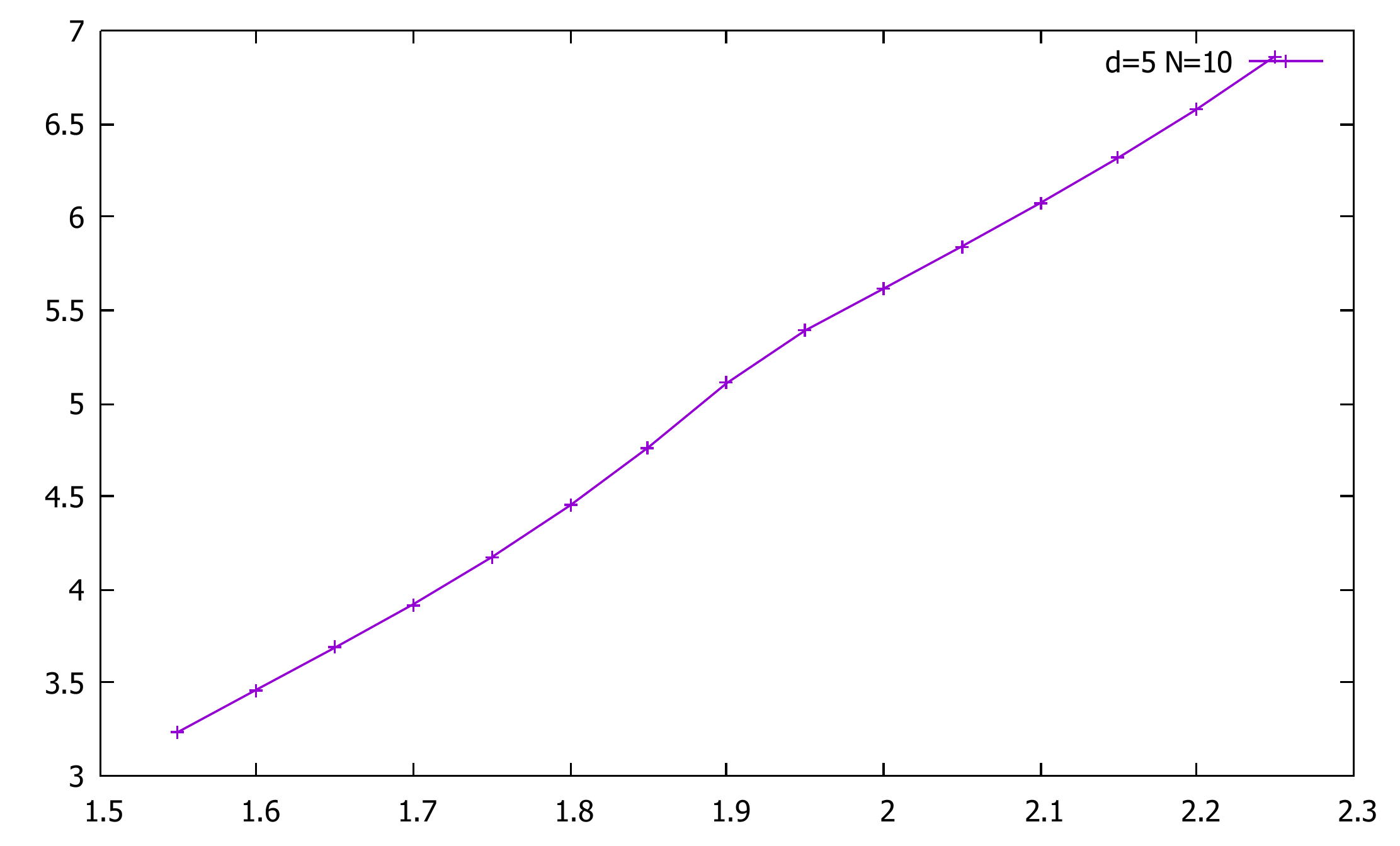}
	\end{center}
	\caption{Unitarity bound in $d=5$ with $N=10$ in the adjoint sector.}
	\label{fig:d5n10}
\end{figure}

\begin{figure}[htbp]
	\begin{center}
		\includegraphics[width=12.0cm,clip]{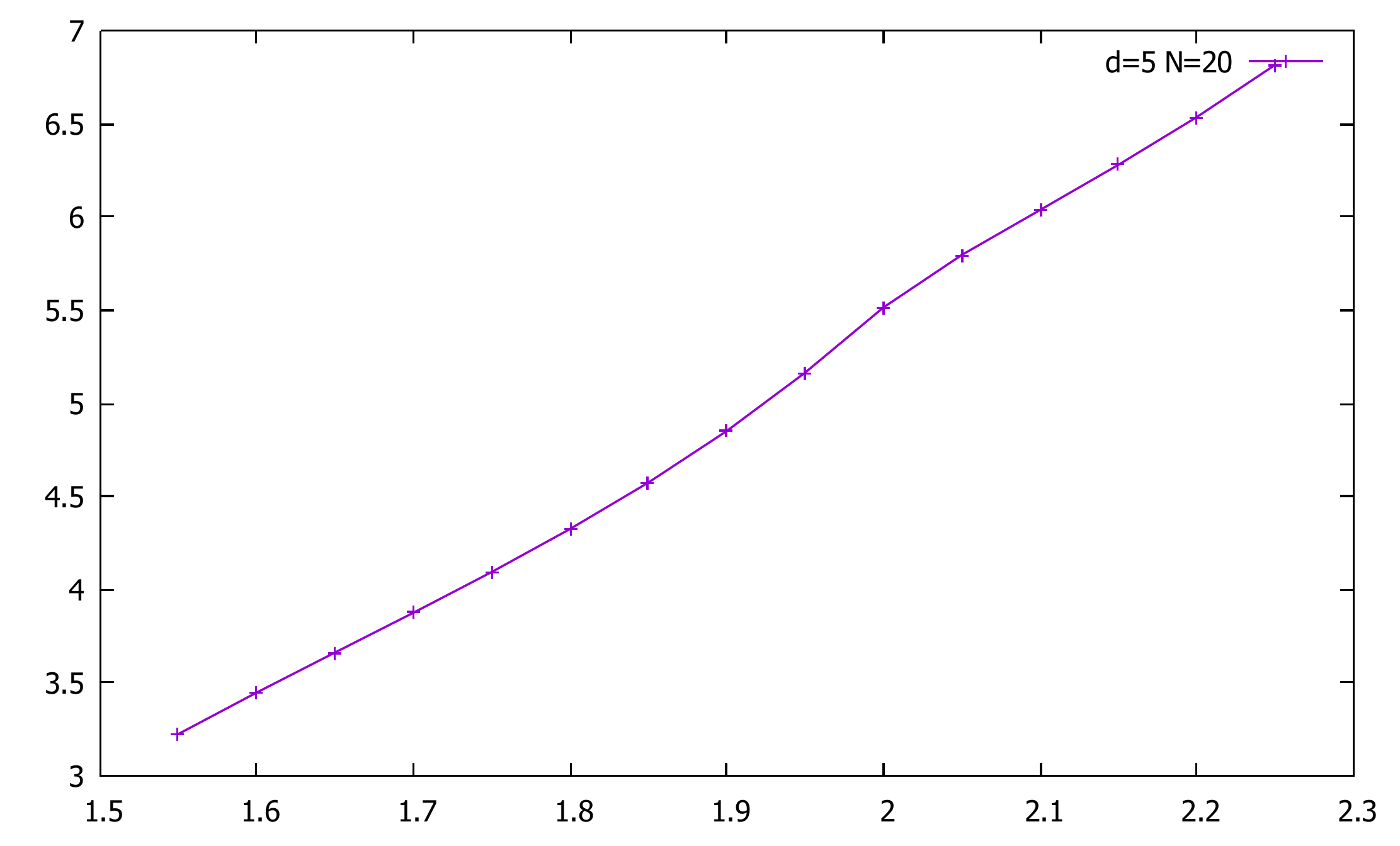}
	\end{center}
	\caption{Unitarity bound in $d=5$ with $N=20$ in the adjoint sector.}
	\label{fig:d5n20}
\end{figure}

\begin{figure}[htbp]
	\begin{center}
		\includegraphics[width=12.0cm,clip]{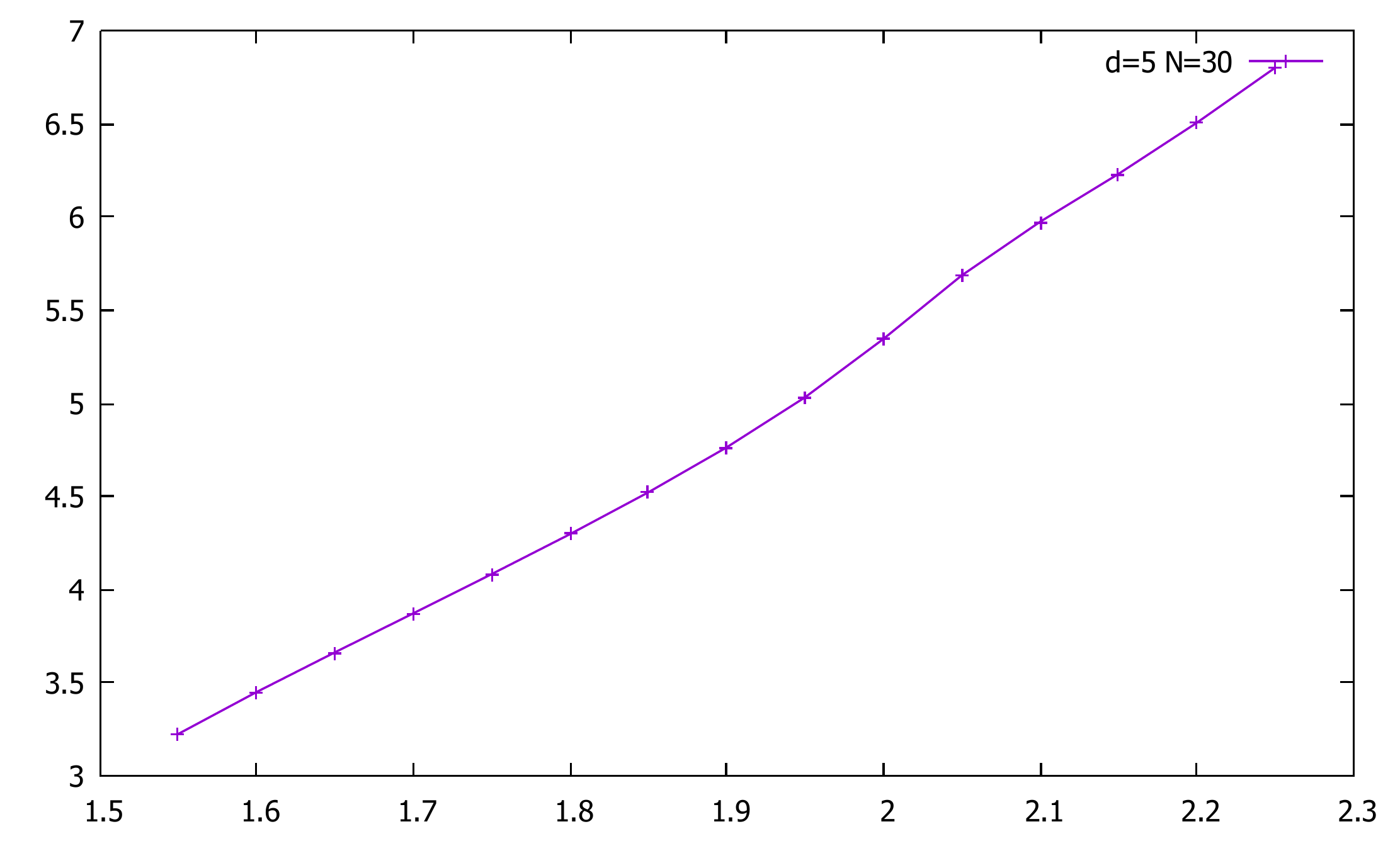}
	\end{center}
	\caption{Unitarity bound in $d=5$ with $N=30$ in the adjoint sector.}
	\label{fig:d5n30}
\end{figure}

\begin{figure}[htbp]
	\begin{center}
		\includegraphics[width=12.0cm,clip]{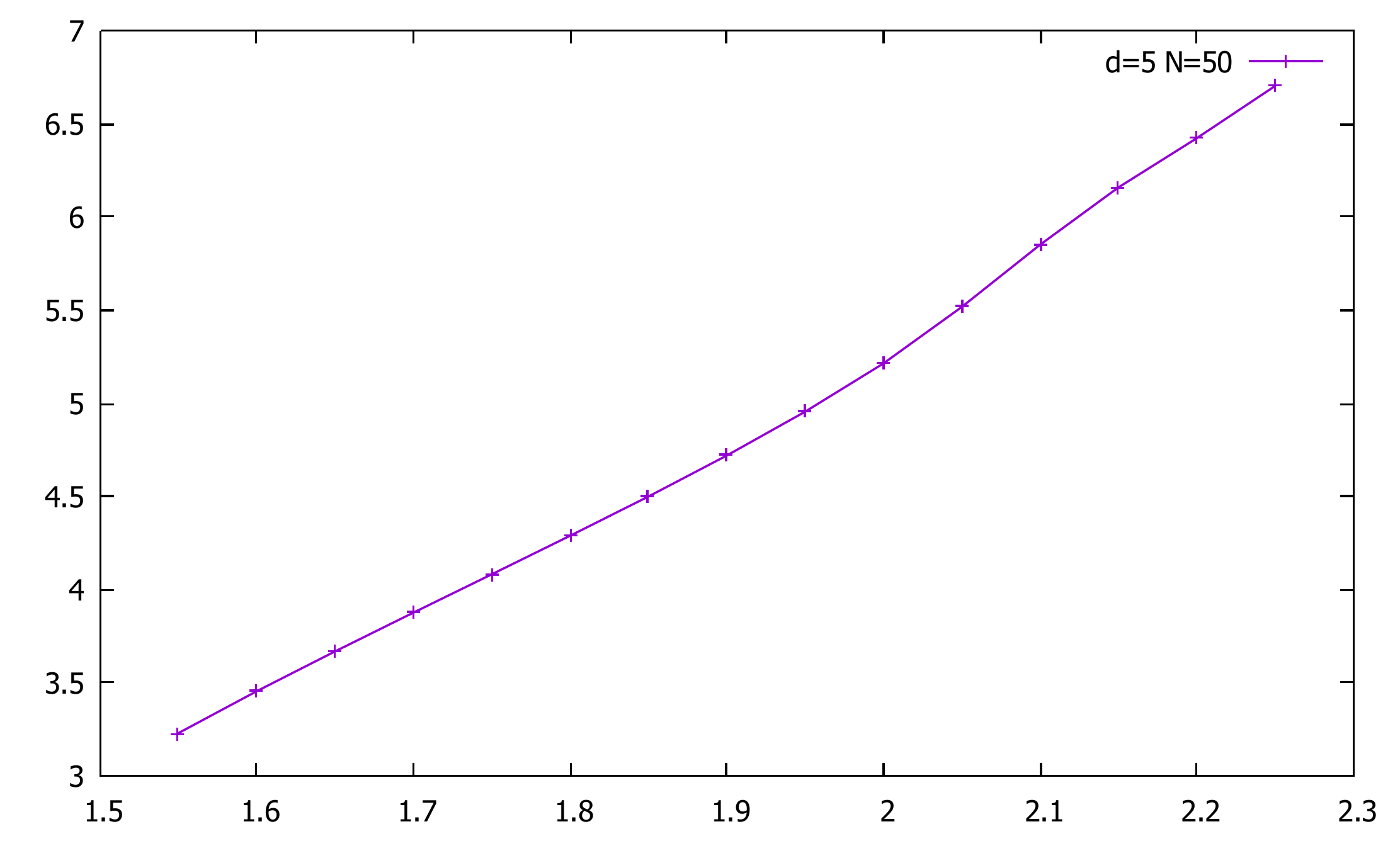}
	\end{center}
	\caption{Unitarity bound in $d=5$ with $N=50$ in the adjoint sector.}
	\label{fig:d5n50}
\end{figure}

\begin{figure}[htbp]
	\begin{center}
		\includegraphics[width=12.0cm,clip]{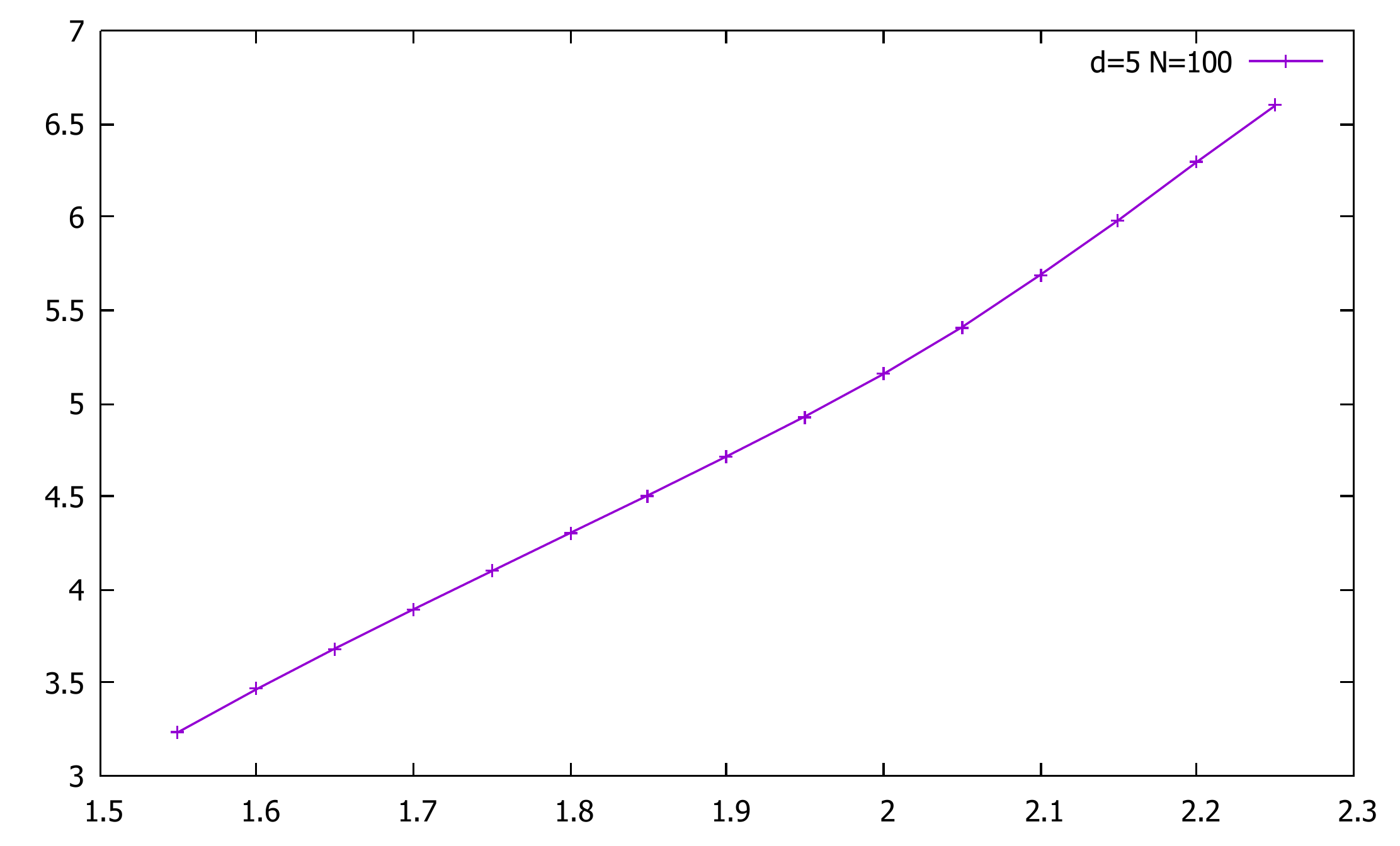}
	\end{center}
	\caption{Unitarity bound in $d=5$ with $N=100$ in the adjoint sector.}
	\label{fig:d5n100}
\end{figure}

In Fig \ref{fig:d4n4}-\ref{fig:d5n100}, we next show the unitarity bound of the scalar operator in the adjoint  representation by changing $N$ in $d=4$ and $d=5$. The case in $d\ge6$ seems uninteresting because as far as we have checked, the features never show up even if we change $N$. Both in $d=4$ and $d=5$, we see that there is a window in which there is a non-trivial feature in the unitarity bound.  The window is $N<N_*$, where $N_* \sim 15 $ in $d=4$ and $N_: \sim 20$ in $d=5$. We further note that $N=4$ does not necessarily gives more eminent features (i.e. sharper kinks) than $N=6$.

We see that the kinks are more or less located when the unitarity bound hits the space-time dimension $d$ even if we change $N$, but the way the features disappear when $N>N_*$ may indicate that the location could be slightly off toward $N \to N_*$ and then eventually disappears, or the dislocation of the kink from $\Delta = d$ may coincidentally indicate the disappearance of the features. 
We need more precise data (e.g. larger search space $\Lambda$ with more sample points) to confirm either of which is realized.

As mentioned at the end of the previous section, one may be able to read the entire conformal data at the boundary of the unitarity bound. Here, we focus on the low-lying spectrum pf conformal dimensions at the kinks in $d=4$ and $d=5$ with $N=6$. We first note that the both spectrum contains a spin two operator with the conformal dimension dimension $\Delta = d$ in the singlet representation of $SU(N)$, which is identified as the energy-momentum tensor. We also have a spin one operator with the conformal dimension $\Delta= d-1$ in the adjoint representation, which is identified as the $SU(N)$ global current. 

In table \ref{table:4}, we show the spectrum of the other scalar operators. It is curious to observe that there are operators in $(A\bar{A})$ representation whose scaling dimension is very close to the unitarity bound of $\Delta \ge \frac{d-2}{2}$. In addition, we find that there is no other relevant operator that is singlet under $SU(N)$ symmetry, so the putative conformal fixed point has only one relevant singlet operator whose conformal dimension is given by $\Delta_1$.

\begin{table}[htbp]
	\resizebox{11cm}{!}{
		\begin{tabular}{|c||c|c|c|c|c|}
			\hline
			&$\Delta_{\Phi}$ & $\Delta_1 $ & $\Delta_{(\mathrm{S\bar{S}})}$ & $\Delta_{(\mathrm{A\bar{A}})}$ & $\Delta_\mathrm{Adj}'$  \\ \hline \hline
			$d=4$ $N=6$ &1.25 & 1.98 & 2.60 & 1.06 & 4.89 \\ \hline 
			$d=5$ $N=6$ &1.85 & 2.49 & 3.66 & 1.56 & 5.66 \\ \hline
	\end{tabular}}
	\caption{The low-lying spectrum read off around the kinks in Fig \ref{fig:d4n6} and Fig \ref{fig:d5n6}. Note that $\Delta_\mathrm{Adj}'$ in the adjoint representation here denotes the value for the second lowest-lying operator. The lowest operator in the adjoint representation has (approximately) $\Delta_{\mathrm{Adj}}=d$ from the location of the kinks.}
	\label{table:4}
\end{table}

\section{Discussions}
We have observed interesting features in the unitarity bound of higher dimensional conformal field theories ($d\ge 4$ in particular) with the $SU(N)$ global symmetry. As we mentioned in the introduction, we have empirical evidence that these features may represent the actual unitary conformal field theories.\footnote{This may have counterexamples. For example, we have  such identifications of the Wilson-Fisher fixed points in $d=4-\epsilon$ dimensions in \cite{El-Showk:2013nia}, but it turned out that these theories are non-unitary \cite{Hogervorst:2015akt} for non-integer $\epsilon$.}
 While this hypothesis of ``kinks = conformal field theories" is not rigorously established, it is encouraging to find interesting features in the unitarity bound from the conformal bootstrap equations.

Assuming that such unitary conformal field theories exist, what would be their origin? The most naive idea to realize a conformal field theory with scalar operators in the adjoint representations of $SU(N)$ global symmetry is to consider matrix theories, where $\Phi$ is identified with a constitutive Hermitian matrix scalar field with the Lagrangian $L = \int d^dx \mathrm{Tr} ( \partial_\mu \Phi \partial^\mu \Phi + V(\Phi))$. However, with the $\mathbb{Z}_2$ symmetry $\Phi \to -\Phi$, the upper critical dimension is believed to be four in these theories, so the simplest idea may not correspond to the putative conformal field theories that we find.\footnote{We do not claim that such a realization is not possible. For example, one may use the qubic interactions and find perturbative fixed points around $d=6$ as discussed in \cite{Fei:2014yja} in the context of vector models.}

The other possible Lagrangian realization is based on gauge theories. Suppose we have $SU(N_c)$ gauge theories with $N_f$ Dirac fermions in the fundamental representation. The theories possess $SU(N_f) \in SU(N_f) \times SU(N_f)$ global symmetry and the gauge invariant ``meson" operators $\bar{\psi}_{\bar{i}}\psi_j$ are in the adjoint representation of the diagonal $SU(N)$. We might wonder the role of the extra non-diagonal flavor symmetry in the conformal bootstrap, but we note that the unitarity bound assuming $SU(N_f) \times SU(N_f)$ is very similar to the one that we have obtained in this paper \cite{Nakayama:2016knq}.

In such realizations, the ``conformal window" we have observed should correspond to the conformal window for the number of flavors in gauge theories. Of course, we have nothing to say about the number of colors, so the interpretation must be tantalizing at best. One may alternatively consider $SU(2)$ gauge theories, and we may study the baryon-like object $\psi_i \psi_j$ , which is in the symmetric representation of $SU(N_f)$. However, since $2$ and $\bar{2}$ is same in $SU(2)$, there is no real distinction between baryons and mesons in $SU(2)$, so our bound actually applies there as well. It is therefore possible that our kinks may be realized by baryon-like operators in $SU(2)$ gauge theories.
We, however, stress that apart from these trivial group theoretic observations, we do not know any further supporting evidence for or against if our kinks have anything to do with the gauge theory realizations of conformal field theories with $SU(N)$ symmetry.

We close this discussion with a historical remark. As far as the author knows, the   discovery of non-trivial kinks in $d=4$ conformal bootstrap program was first done by T.~Ohtsuki in his analysis of four-point functions among ``symmetric representations" of $SU(N)$  in the adjoint sector precisely for the purpose of identifying them as   four-point functions of baryon operators in $SU(2)$ gauge theories. For the reasons discussed above, as a $SU(2)$ gauge theory, the study of the $SU(N)$ adjoint sector done in this paper is more or less equivalent to the one   in the symmetric representation. 
Indeed, his analysis of the $SU(N)$ symmetric representations, which is numerically a little more complicated that ours, also gave the kinks which are very similar to ours discussed in this paper \cite{unpublished}.\footnote{During the preparation of this work, A.~Vichi informed us that he has been pursuing the similar directions in the study of the conformal bootstrap of the symmetric representation of $SU(N)$ global symmetry \cite{Vic}.}

\section*{Acknowledgements}
The author would like to thank T.~Ohtsuki for his dedications to conformal bootstrap. Without his help, this work would have never appeared. He would also like to thank A.~Vichi for the correspondence. This work is supported in part by Rikkyo University Special Fund for Research.


\appendix
\section{Kinks in the other sectors}
In the main part of this paper, we only discussed the unitarity bound in the adjoint sector. It is, however, an interesting question if there are any other features in the unitarity bound in the other sectors of our $\Phi \times \Phi$ OPE. Indeed there are, and in this appendix, we report some of them.

\begin{figure}[htbp]
	\begin{center}
		\includegraphics[width=12.0cm,clip]{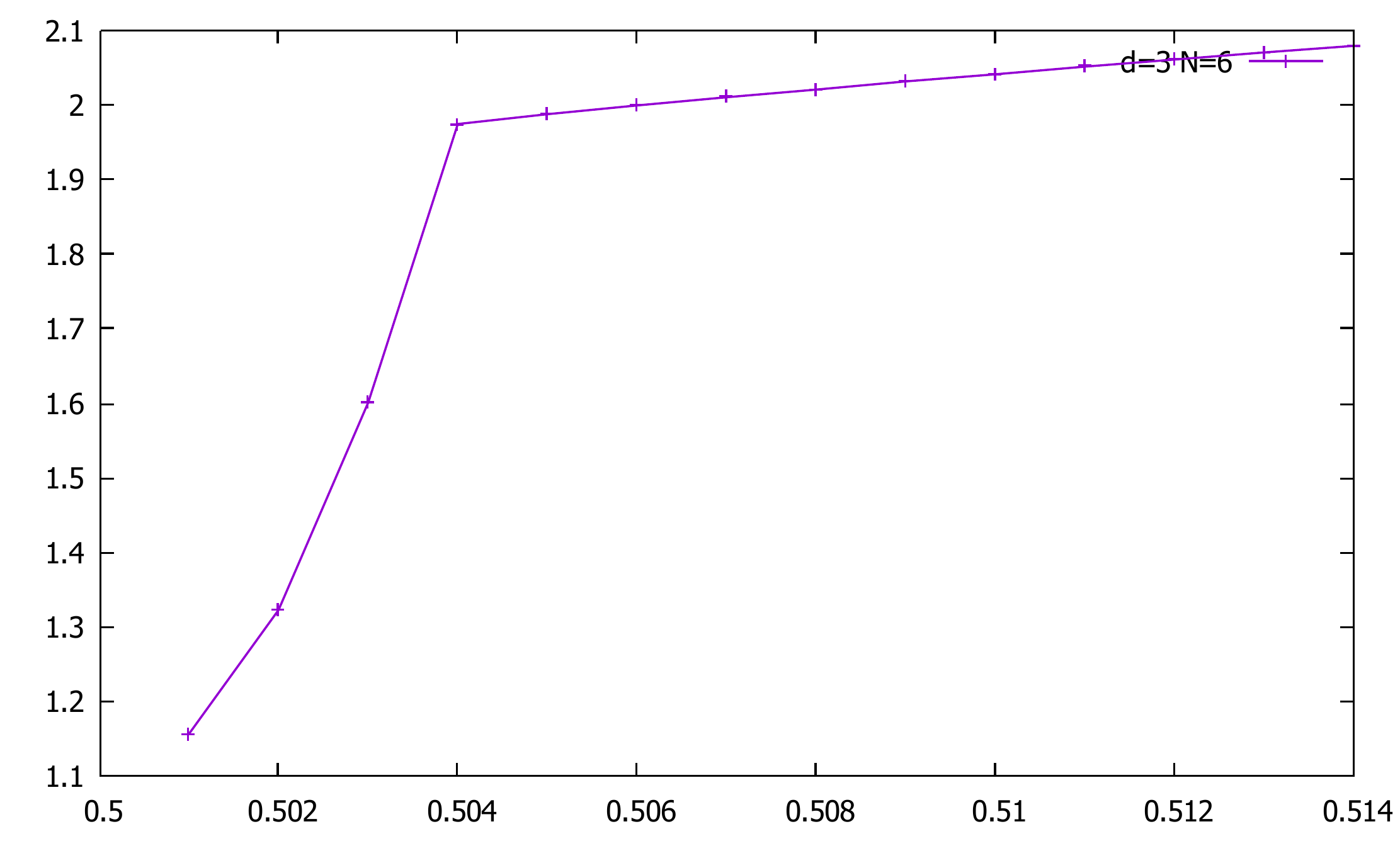}
	\end{center}
	\caption{Unitarity bound in $d=3$ with $N=6$ in the singlet sector.}
	\label{fig:d3n6singlet}
\end{figure}

First of all, we find that the scalar operators in the singlet representation has kinks in $d=3$ dimensions that are (almost) identical to the ones in the $O(\tilde{N})$ model studied in \cite{Kos:2013tga}. This makes sense because we cannot exclude the conformal field theories realized by $O(\tilde{N})$ vector models by identifying $SU(N)$ adjoint  operator as $O(\tilde{N})$ fundamental where $\tilde{N} = N^2-1$. This phenomenon is known as a symmetry enhancement in the singlet sector and it has been reported in various places with different symmetries. We present our numerical results with $d=3$ and $N=6$ in Fig \ref{fig:d3n6singlet}.

\begin{figure}[htbp]
	\begin{center}
		\includegraphics[width=12.0cm,clip]{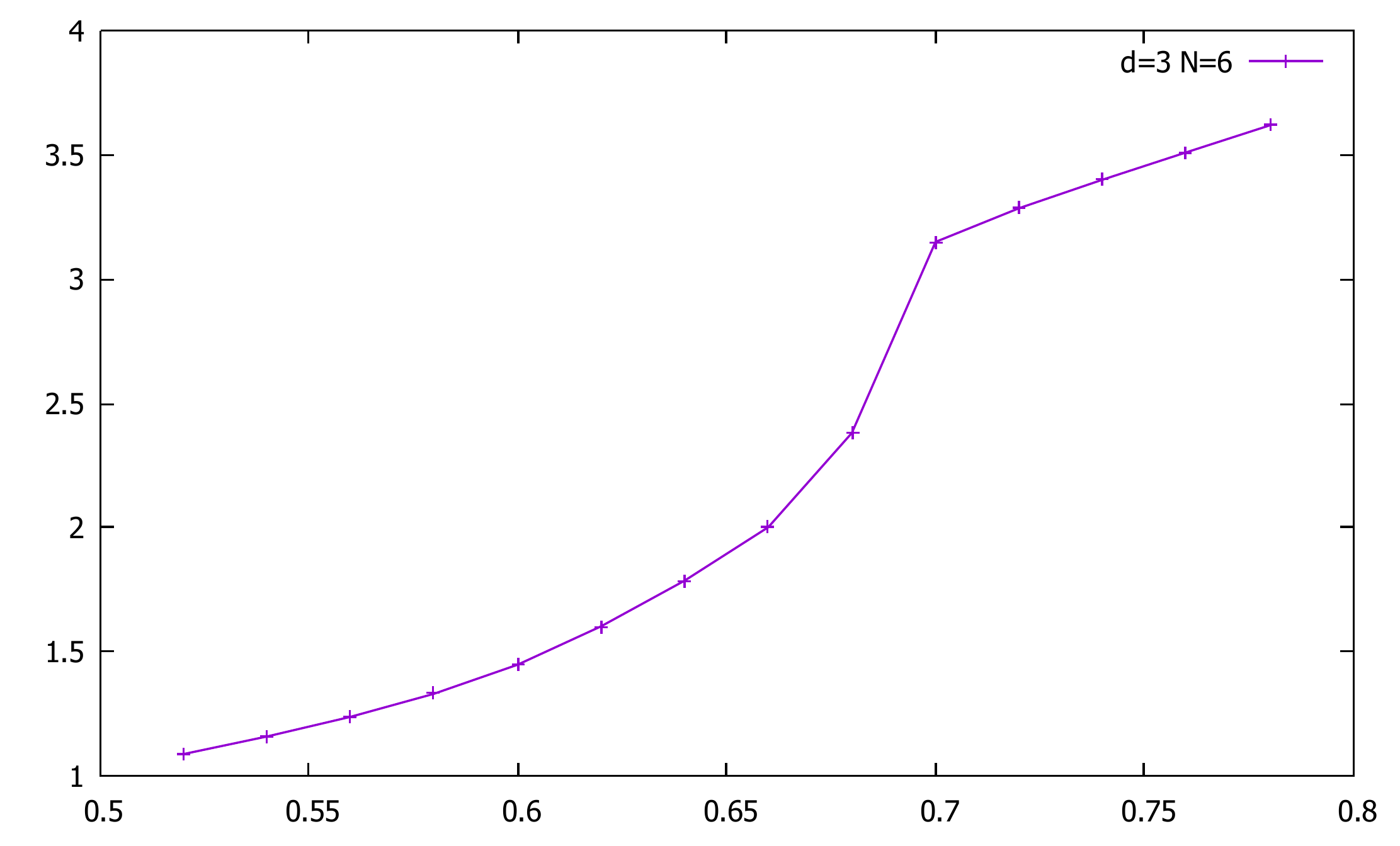}
	\end{center}
	\caption{Unitarity bound in $d=3$ with $N=6$ in the $\mathrm{(A\bar{A})}$  sector.}
	\label{fig:d3n6aas}
\end{figure}

\begin{figure}[htbp]
	\begin{center}
		\includegraphics[width=12.0cm,clip]{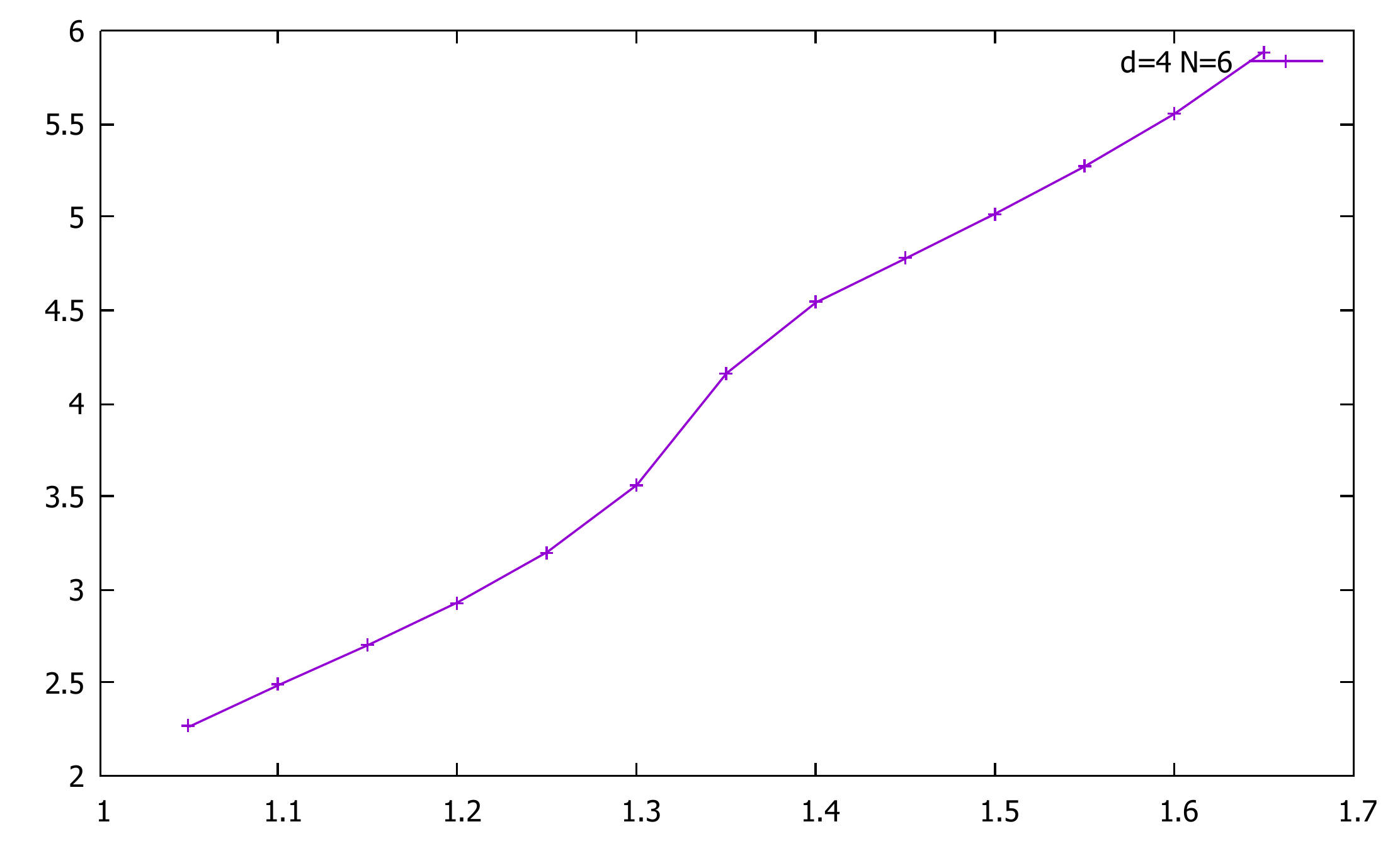}
	\end{center}
	\caption{Unitarity bound in $d=4$ with $N=6$ in the $\mathrm{(A\bar{A})}$  sector.}
	\label{fig:d4n6aas}
\end{figure}

\begin{figure}[htbp]
	\begin{center}
		\includegraphics[width=12.0cm,clip]{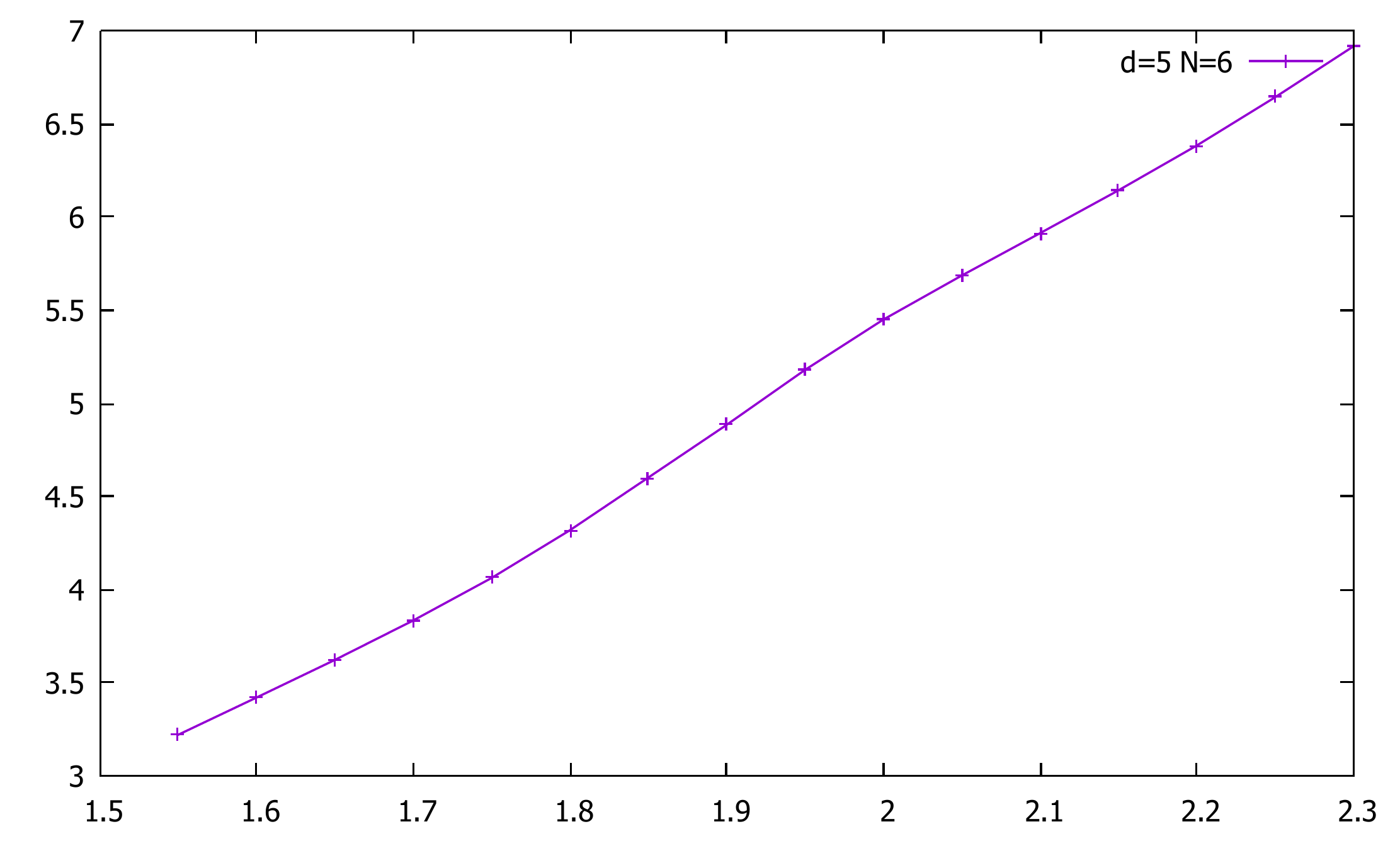}
	\end{center}
	\caption{Unitarity bound in $d=5$ with $N=6$ in the $\mathrm{(A\bar{A})}$  sector.}
	\label{fig:d5n6aas}
\end{figure}

\begin{figure}[htbp]
	\begin{center}
		\includegraphics[width=12.0cm,clip]{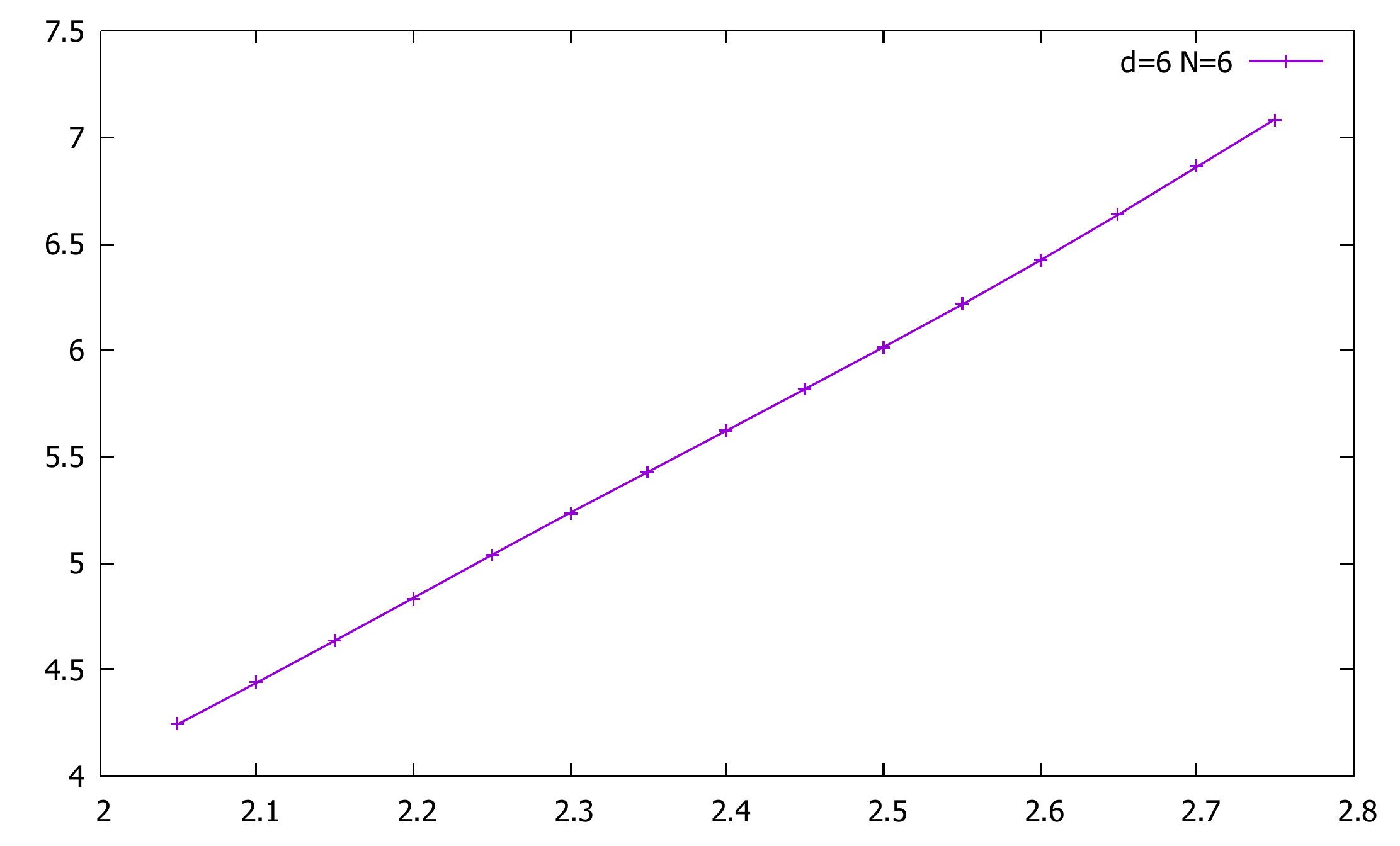}
	\end{center}
	\caption{Unitarity bound in $d=6$ with $N=6$ in the $\mathrm{(A\bar{A})}$  sector.}
	\label{fig:d6n6aas}
\end{figure}

There are other kinks that are yet to be identified. As first observed by Iha et al \cite{Iha:2016ppj}, there is an interesting feature even in $d=4$ if we consider the $\mathrm{(A\bar{A})}$ sector. In Fig \ref{fig:d3n6aas}-\ref{fig:d6n6aas} we report the unitarity bound in the $\mathrm{(A\bar{A})}$ sector with $N=6$ in $d=3,4,5,6$ dimensions. We see the most eminent kink in $d=3$ and the kink becomes less sharp in higher dimensions. In $d=6$, it disappears completely.

Note that the properties of the putative CFTs realized at these kinks are distinct from those discussed in the main text. Not only the locations of kinks as a function of $\Delta_{\Phi}$ are different, but we recall that $\Delta_{\mathrm{(A\bar{A})}}$ there was close to the unitarity bound. Here, this is the conformal dimension of the operator that are bounded, and it can be as high as  $\Delta_{\mathrm{(A\bar{A})}} \sim d$ around the kinks.

In the main text, we presented the possibility that the observed kinks may be realized as gauge theories with $SU(N)$ flavor symmetry. If the kinks presented in this appendix also correspond to conformal field theories from such gauge theories, we have at least two quite distinct fixed points. This was not theoretically unlikely as discussed in the context of the Landau-Ginzburg model in \cite{Nakayama:2014lva}\cite{Nakayama:2014sba}, in which case we had two different fixed points with respect to quartic coupling constants.
Whether we can give such an interpretation to the kinks here is an open question.

\end{document}